\documentclass[twocolumn,pra,showpacs,showkeys]{revtex4}
\usepackage{amsfonts}
\usepackage{amsmath}
\usepackage{amssymb}
\usepackage{graphicx}
\usepackage{rotating}

\providecommand{\U}[1]{\protect\rule{.1in}{.1in}}
\providecommand{\U}[1]{\protect\rule{.1in}{.1in}}
\providecommand{\U}[1]{\protect\rule{.1in}{.1in}}

\begin{document}

\preprint{}
\title{Effect of impurities on nuclear fusion}
\author{P\'{e}ter K\'{a}lm\'{a}n}
\author{Tam\'{a}s Keszthelyi}
\affiliation{Budapest University of Technology and Economics, Institute of Physics,
Budafoki \'{u}t 8. F., H-1521 Budapest, Hungary\ }

\begin{abstract}
Modification of nuclear reactions due to impurities in plasma is
investigated. The hindering effect of Coulomb repulsion between reacting
particles, that is effective in direct reactions, can practically disappear
if Coulomb interaction of the reacting particles with impurities embedded in
plasma is taken into account. The change of the wavefunction of reacting
particles in nuclear range due to their Coulomb interaction with impurity is
determined using standard time independent perturbation calculation of
quantum mechanics. The result can be interpreted as if a slow, quasi-free
particle (e.g. a proton) were pushed by a heavy, assisting particle
(impurity) of the surroundings and can get (virtually) such a great
magnitude of momentum which significantly increases the nuclear contact
probability density and also the probability of its capture by an other
nucleus. As a sample reaction the process, called impurity assisted nuclear $%
pd$ reaction is investigated and the rate and power densities produced by
the reaction are numerically calculated. With the aid of astrophysical
factors the rate and power densities of the impurity assisted $%
d(d,n)_{2}^{3}He$, $d(d,p)t$, $d(t,n)_{2}^{4}He$, $_{2}^{3}He(d,p)_{2}^{4}He$%
, $_{3}^{6}Li(p,\alpha )_{2}^{3}He$, $_{3}^{7}Li(p,\alpha )_{2}^{4}He$, $%
_{4}^{9}Be(p,\alpha )_{3}^{6}Li$, $_{4}^{9}Be(p,d)_{4}^{8}Be$, $%
_{4}^{9}Be(\alpha ,n)_{6}^{12}C$, $_{5}^{10}B(p,\alpha )_{4}^{7}Be$ and $%
_{5}^{11}B(p,\alpha )_{4}^{8}Be$ reactions are also estimated. The affect of
plasma-wall interaction on the process is also considered. A partial survey
of impurity assisted nuclear reactions which may have practical importance
in energy production is also presented.
\end{abstract}

\volumenumber{number}
\issuenumber{number}
\eid{identifier}
\date[Date text]{date}
\received[Received text]{date}
\revised[Revised text]{date}
\accepted[Accepted text]{date}
\published[Published text]{date}
\startpage{1}
\endpage{}
\maketitle

\section{Introduction}

Nuclear fusion reactors need to be heated to very high temperature to overcome the Coulomb repulsion between nuclei to fuse. 
\cite{McCracken}. (Mathematically it is manifested in the exponential energy
dependent factor in the cross section of fusion reactions [see $\left( \ref%
{sigma}\right) $]. For details see Section II.)

The effect of surroundings on nuclear fusion rate in astrophysical condensed
and dense laboratory plasmas has been extensively studied in the case of usual
nuclear reactions. In tenuous plasmas the effect of spectator nuclei and
electrons (the environment) on the Gamow-rate of reacting nuclei which are
assumed to interact with bare Coulomb potential is negligible \cite{Ichimaru}%
. Moreover in some (e.g. tokamak-like) devices the presence of impurities
during the heat up and working periods is undesirable because of high
loss power generated by them \cite{Dolan}.

In this paper it will be shown however, that spectator nuclei can
significantly enhance nuclear processes and allow new types of reactions.

We are going to investigate these processes of new type that can take place
due to impurities and their effect on nuclear fusion devices. We focus our
attention on Coulomb interaction between fuel nuclei and environment, namely
on consequences of interactions with impurities which can activate new type
of reactions of cross section of considerable magnitude which may change the
condition of necessary plasma temperature and what is more remarkable, the mechanism
found does not need plasma state at all.

We investigate the%
\begin{equation}
\text{ }_{z_{1}}^{A_{1}}V+p+\text{ }_{z_{3}}^{A_{3}}X\rightarrow \text{ }%
_{z_{1}}^{A_{1}}V^{\prime }+\text{ }_{z_{3}+1}^{A_{3}+1}Y+\Delta
\label{Reaction 1}
\end{equation}%
process called impurity assisted proton capture, a process among atoms or
atomic ions containing $_{z_{1}}^{A_{1}}V$ nuclei (e.g. $Xe$) and protons or
hydrogen atoms and ions or atoms of nuclei $_{z_{3}}^{A_{3}}X$ (e.g.
deuterons) that are initially supposed to be in a plasma.
$\Delta $ is the energy of the reaction. In normal case proton
capture may happen in the%
\begin{equation}
p+\text{ }_{z_{3}}^{A_{3}}X\rightarrow \text{ }_{z_{3}+1}^{A_{3}+1}Y+\gamma
+\Delta .  \label{p-capture}
\end{equation}%
reaction where $\gamma $ emission is required by energy and
momentum conservation. Accordingly $\left( \ref{Reaction 1}\right) $ describes a new type of $p$%
-capture. In the usual $p$-capture reaction $\left( \ref%
{p-capture}\right) $\ particles $_{z_{3}+1}^{A_{3}+1}Y$ and $\gamma $ take
away the reaction energy and the reaction is governed by electromagnetic interaction.
 In reaction $\left( \ref{Reaction 1}\right) $ the reaction energy is taken away by particles $%
_{z_{1}}^{A_{1}}V^{\prime }$ and $_{z_{3}+1}^{A_{3}+1}Y$ 
while the reaction is governed by Coulomb as well as strong interactions.

First we pay our attention to the impurity assisted $p+d\rightarrow $ $%
^{3}He $ reaction%
\begin{equation}
_{z_{1}}^{A_{1}}V+p+d\rightarrow \text{ }_{z_{1}}^{A_{1}}V^{\prime }+\text{ }%
_{2}^{3}He+5.493MeV  \label{Reaction 2}
\end{equation}%
in an impurity contaminated plasma, that will be discussed in more detail.

Cross section and rate of process $\left( \ref{Reaction 1}\right) $ to be
considered can be calculated by the rules of time-independent
perturbation calculation of quantum mechanics \cite{Landau}. Our results
indicate that the cross section of process $\left( \ref{Reaction 1}\right) $%
, which is an indirect (second order) reaction, may be essentially higher
with decreasing energy than the cross section of a usual, direct (first
order) reaction since the huge exponential drop in the cross section $\left( %
\ref{sigma}\right) $ with decreasing energy disappears from the cross
section of process $\left( \ref{Reaction 1}\right) $.

The disappearence of the exponential energy dependent factor in $\left( \ref{sigma}%
\right) $ means that due to impurity assisted reactions $\left( \ref{Reaction 1}\right) $ the extra high temperature needed to ignite nuclear fusion in plasma may be appreciably reduced.

\begin{figure}[tbp]
\resizebox{6.0cm}{!}{\includegraphics*{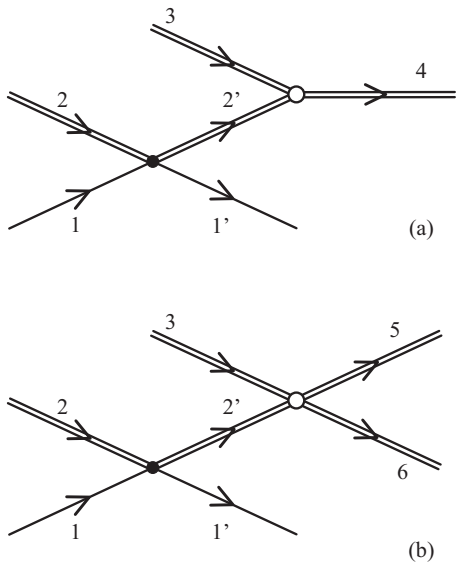}}
\caption{The graphs of impurity assisted nuclear reactions. The single lines
represent (initial (1) and final (1')) impurity particle of the plasma. The
double lines represent free, heavy initial (2) particles (such as $p$, $d$),
their intermediate state (2'), target nuclei (3) and reaction products (4,
5, 6). The filled dot denotes Coulomb-interaction and the open circle
denotes nuclear (strong) interaction. FIG. 1(a) is a capture process and
FIG. 1(b) is a reaction with two fragments.}
\end{figure}

A less precise picture, still abiding the rules of second
order time-independent perturbation calculation of quantum mechanics, can
help to understand the effect with the aid of graphs
 (Fig. 1.) The physics behind the calculation may be
interpreted as follows. The Coulomb interaction between
particles $_{z_{1}}^{A_{1}}V$ and protons mixes (an intermediate) state of
the proton of large momentum to the initially slow protonic state with a
small amplitude while particle $_{z_{1}}^{A_{1}}V$ is recoiled. Thus the
Coulomb interaction pushes the protons (virtually) into an intermediate
state. In this state protons have large enough (virtual) momentum to get
over the Coulomb repulsion of nuclei $_{z_{3}}^{A_{3}}X$ and so they may be 
captured by the nuclei $_{z_{3}}^{A_{3}}X$ due to strong interaction to
create a $_{z_{3}+1}^{A_{3}+1}Y$ nucleus. The particles (impurities) $%
_{z_{1}}^{A_{1}}V$ (initial) and $_{z_{1}}^{A_{1}}V^{\prime }$ (final)
assist the process only.

The virtual momentum of the intermediate state can be determined in the
following way. Energy and momentum conservations determine the wave-vectors $%
\mathbf{k}_{1^{\prime }}$ and $\mathbf{k}_{4}$, of
particles 1' and 4, respectively, as $\mathbf{k}_{1^{\prime }}=-\mathbf{k}%
_{4}$ and $\left\vert \mathbf{k}_{1^{\prime }}\right\vert =\left\vert 
\mathbf{k}_{4}\right\vert =k_{0}$ with $k_{0}=\hbar ^{-1}\sqrt{%
2m_{0}a_{14}\Delta }$. Here $m_{0}c^{2}=931.494$ $MeV$ is the atomic energy
unit and $a_{14}$ is determined by $\left( \ref{ajk}\right) $. (The initial momenta and kinetic energies are assumed to be negligible.) Because
of momentum conservation in Coulomb scattering of plane waves the wave
vector $\mathbf{k}_{2^{\prime }}$ of particle 2' is determined as $\mathbf{k}%
_{2^{\prime }}=-\mathbf{k}_{1^{\prime }}$, i.e. $\left\vert \mathbf{k}%
_{2^{\prime }}\right\vert =k_{0}$ too. Consequently $\left\vert \mathbf{k}%
_{2^{\prime }}\right\vert $ is large enough for particle 2' to effectively
overcome the Coulomb repulsion. (For the details of the rigorous calculations see in Section III.)

A generalization of $\left( \ref{Reaction 1}\right) $ can be the reaction%
\begin{equation}
_{z_{1}}^{A_{1}}V+\text{ }_{z_{2}}^{A_{2}}w+\text{ }_{z_{3}}^{A_{3}}X%
\rightarrow \text{ }_{z_{1}}^{A_{1}}V^{\prime }+\text{ }%
_{z_{3}+z_{2}}^{A_{3}+A_{2}}Y+\Delta  \label{Reaction 3}
\end{equation}%
that will be briefly discussed to draw conclusions as to the possible
modification of appropriate fuels of nuclear fusion reactors by impurity
assisted reactions.

As a further generalization the reaction%
\begin{equation}
_{z_{1}}^{A_{1}}V+\text{ }_{z_{2}}^{A_{2}}w+\text{ }_{z_{3}}^{A_{3}}X%
\rightarrow \text{ }_{z_{1}}^{A_{1}}V^{\prime }+\text{ }_{z_{4}}^{A_{4}}Y+%
\text{ }_{z_{5}}^{A_{5}}W+\Delta  \label{Reaction 4}
\end{equation}%
with two final fragments is also considered. The impurity
assisted $d(d,n)_{2}^{3}He$, $d(d,p)t$, $d(t,n)_{2}^{4}He$ and $%
_{2}^{3}He(d,p)_{2}^{4}He$ reactions are numerically investigated and their rate and
power densities are also determined.

In Section II. the essential role of the Coulomb factor is discussed.
Section III. is devoted to the discussion of the model. The change of the
wavefunction in the nuclear range due to the impurity is determined. The
transition probability per unit time, cross section, rate and power
densities of impurity assisted $p+d\rightarrow $ $^{3}He$ reaction, which is
the simplest impurity assisted proton capture reaction in an atomic-atom
ionic gas mix, are given. The cross section of impurity assisted reactions
with two final fragments are determined and the affect of plasma-wall
interaction on the process is also considered. In Section IV. the rate and
power densities in a $p-d-Xe$ atomic atom-ionic gas mix are calculated
numerically. Section V. is a partial overview of some other impurity
assisted nuclear reactions and gives account of the estimated power densities
of the impurity assisted $d(d,n)_{2}^{3}He$, $d(d,p)t$, $d(t,n)_{2}^{4}He$, $%
_{2}^{3}He(d,p)_{2}^{4}He$, $_{3}^{6}Li(p,\alpha )_{2}^{3}He$, $%
_{3}^{7}Li(p,\alpha )_{2}^{4}He$, $_{4}^{9}Be(p,\alpha )_{3}^{6}Li$, $%
_{4}^{9}Be(p,d)_{4}^{8}Be$, $_{4}^{9}Be(\alpha ,n)_{6}^{12}C$, $%
_{5}^{10}B(p,\alpha )_{4}^{7}Be$ and $_{5}^{11}B(p,\alpha )_{4}^{8}Be$ 
reactions. Section VI. is a Summary.

\section{Coulomb factor}

The cross section $\left( \sigma \right) $ of usual fusion reactions between
particles $2$ and $3$ reads as \cite{Angulo} 
\begin{equation}
\sigma \left( E\right) =S\left( E\right) \exp \left[ -2\pi \eta _{23}\left(
E_{l}\right) \right] /E,  \label{sigma}
\end{equation}%
where $E$ is the energy taken in the center of mass $\left( CM\right) $
coordinate system.%
\begin{equation}
\eta _{23}=z_{2}z_{3}\alpha _{f}\frac{a_{23}m_{0}c}{\hbar \left\vert \mathbf{%
k}\right\vert }=z_{2}z_{3}\alpha _{f}\sqrt{a_{23}\frac{m_{0}c^{2}}{2E}}
\label{etajk}
\end{equation}%
is the Sommerfeld parameter. Here $\mathbf{k}$ is the wave number vector of
the particles $2$ and $3$ in their relative motion, $\hbar $ is the reduced
Planck-constant, $c$ is the velocity of light in vacuum and%
\begin{equation}
a_{jk}=\frac{A_{j}A_{k}}{A_{j}+A_{k}}  \label{ajk}
\end{equation}%
is the reduced mass number of particles $j$ and $k$ of mass numbers $A_{j}$
and $A_{k}$ and rest masses $m_{j}=A_{j}m_{0}$, $m_{k}=A_{k}m_{0}$. $%
m_{0}c^{2}=931.494$ $MeV$ is the atomic energy unit, $\alpha _{f}$ is the
fine structure constant. The cross section $\left( \ref{sigma}\right) $ can
be\ derived applying an approximate form%
\begin{equation}
\varphi _{Cb,a}(\mathbf{r})=f_{23}(k\left[ E\right] )\exp (i\mathbf{k\cdot
r)/}\sqrt{V}  \label{Cbapp}
\end{equation}%
of the Coulomb solution $\varphi _{Cb}(\mathbf{r})$ \cite{Alder} valid in
the nuclear volume that produces the same $\left\vert \varphi (\mathbf{0}%
)_{Cb,a}\right\vert ^{2}=f_{23}^{2}/V$ contact probability density as $%
\varphi _{Cb}(\mathbf{r})$. Here $\mathbf{r}=\mathbf{r}_{2}-\mathbf{r}_{3}$
is the relative coordinate of particles $2$ and $3$ of coordinate $\mathbf{r}%
_{2}$ and $\mathbf{r}_{3}$, and $f_{23}(k)$ is the Coulomb factor 
\begin{equation}
f_{23}=\left\vert e^{-\pi \eta _{23}/2}\Gamma (1+i\eta _{23})\right\vert =%
\sqrt{\frac{2\pi \eta _{23}}{\exp \left( 2\pi \eta _{23}\right) -1}}
\label{Fjk}
\end{equation}%
corresponding to particles $2$ and $3$.

The cross section is proportional to $f_{23}^{2}$ and one can show that the
exponential factor in $\left( \ref{sigma}\right) $ comes from $f_{23}^{2}(E)$%
. Thus the smallness of rate at low energies is the consequence of $%
f_{23}^{2}(E)$ becoming very small at lower energies. So the magnitude of
the Coulomb factor $f_{23}(E)$ is crucial from the point of view of
magnitude of the cross section.

\section{Model of impurity assisted nuclear reactions in an atomic-atom
ionic gas mix}

\subsection{Change of wavefunction in nuclear range}

We focus on the modification of nuclear reactions in a plasma.
At first capture processes are dealt with. It is supposed that all components
are in atomic, atom-ionic or fully-ionized state while the necessary number
of electrons required for electric neutrality are also present.

Let us take three screened charged particles of rest masses $m_{j}$ ($%
j=1,2,3$). Particles are heavy and have nuclear charges $z_{j}e$ with $e$
the elementary charge and $z_{j}$ the charge number. (The coupling strength $%
e^{2}=\alpha _{f}\hbar c$.) The total Hamiltonian which describes this
3-body system is 
\begin{equation}
H_{tot}=H_{1}+H_{23}+V_{Cb}(1,2)+V_{Cb}(1,3),  \label{Htot}
\end{equation}%
where $H_{1}=H_{kin,1}$ is the Hamiltonian of particle $1$ which is
considered to be free ($H_{kin,j}$ denotes the kinetic Hamiltonian of
particle $j$),%
\begin{equation}
H_{23}=H_{kin,2}+H_{kin,3}+V_{Cb}(1,3)  \label{H23}
\end{equation}%
is the Hamiltonian of particles $2$ and $3$. Their nuclear reaction will be
discussed later. Here and in $\left( \ref{Htot}\right) $ $V_{Cb}(j,k)$
denotes screened Coulomb interaction between particles $j$ and $k$ with
screening parameter $q_{sc,jk}$ and of form in the coordinate representation%
\begin{equation}
V_{Cb}\left( j,k\right) =\frac{z_{j}z_{k}e^{2}}{2\pi ^{2}}\int \frac{\exp (i%
\mathbf{qr}_{jk}\mathbf{)}}{q^{2}+q_{sc,jk}^{2}}d\mathbf{q,}  \label{Vcb1}
\end{equation}%
where $\mathbf{r}_{jk}=\mathbf{r}_{j}-\mathbf{r}_{k}$ is the relative
coordinate between particles $j$ and $k$ of coordinate $\mathbf{r}_{j}$ and $%
\mathbf{r}_{k}$. $H_{kin,2}$ and $H_{kin,3}$ are the kinetic Hamiltonians of
particles $2$ and $3$.

It is supposed that stationary solutions $\left\vert 1\right\rangle $ and $%
\left\vert 2,3\right\rangle _{sc}$ of energy eigenvalues $E_{1}\emph{\ }$and 
$E_{23}$ of the stationary Schr\"{o}dinger equations $H_{1}\left\vert
1\right\rangle =E_{1}\left\vert 1\right\rangle $ with $E_{1}$ the kinetic
energy of particle $1$ and $H_{23}\left\vert 2,3\right\rangle
_{sc}=E_{23}\left\vert 2,3\right\rangle _{sc}$ with $E_{23}=E_{CM}+E_{rel}$
are known. Here $E_{CM}$ and $E_{rel}$ are the energies attached to the
center of mass $\left( CM\right) $ and relative motions of particles $2$ and 
$3$. Thus $H_{tot}$ can be written as $H_{tot}=H_{0}+H_{Int}$ with $%
H_{0}=H_{1}+H_{23}$ as the unperturbed Hamiltonian and%
\begin{equation}
H_{Int}=V_{Cb}(1,2)+V_{Cb}(1,3)  \label{Hint}
\end{equation}%
as the interaction Hamiltonian (perturbation). The stationary solution $%
\left\vert 1,2,3\right\rangle _{sc}$ of $H_{0}\left\vert 1,2,3\right\rangle
_{sc}=E_{0}\left\vert 1,2,3\right\rangle _{sc}$ with $E_{0}=E_{1}+E_{23}$
can be written as $\left\vert 1,2,3\right\rangle _{sc}=\left\vert
1\right\rangle \left\vert 2,3\right\rangle _{sc}$ which is the direct
product of states $\left\vert 1\right\rangle $ and $\left\vert
2,3\right\rangle _{sc}$. The states $\left\vert 1,2,3\right\rangle _{sc}$
form an orthonormal complete system.

The approximate solution of $H_{tot}\left\vert 1,2,3\right\rangle
_{sc}=E_{0}\left\vert 1,2,3\right\rangle _{sc}$ in the screened case is
obtained with the aid of standard time independent perturbation calculation 
\cite{Landau} and the first order approximation is expanded in terms of the
complete system. The terms which differ from the initial state and which are
considered to be intermediate from the point of view of the next step of
perturbation calculation taking into account strong interaction will be
called intermediate states.

The solution of $H_{23}\left\vert 2,3\right\rangle =E_{23}\left\vert
2,3\right\rangle $ in the unscreened case is known and the coordinate
representation $\langle \mathbf{R,r}\left\vert 2,3\right\rangle =\varphi
_{23}\left( \mathbf{R,r}\right) $ of $\left\vert 2,3\right\rangle $ has the
form%
\begin{equation}
\varphi _{23}\left( \mathbf{R,r}\right) =\frac{e^{i\mathbf{KR}}}{\sqrt{V}}%
\varphi _{Cb}(\mathbf{r}),  \label{solution23}
\end{equation}%
where $\mathbf{K}$, $\mathbf{R}=\left( m_{2}\mathbf{r}_{2}+m_{3}\mathbf{r}%
_{3}\right) /\left( m_{2}+m_{3}\right) $ and $\mathbf{r=r}_{23}$ are wave
vector of the $CM$ motion, $CM$ and relative coordinate of particles $2$ and 
$3$, respectively, $V$ denotes the volume of normalization and 
\begin{equation}
\varphi _{Cb}(\mathbf{r})=e^{i\mathbf{k}\cdot \mathbf{r}}f(\mathbf{k,r})/%
\sqrt{V}  \label{Cb1}
\end{equation}%
is the unscreened Coulomb solution \cite{Alder}. Here $\mathbf{k}$ is the
wave number vector of the particles $2$ and $3$ in their relative motion and 
$f(\mathbf{k},\mathbf{r})=e^{-\pi \eta _{23}/2}\times $ $\Gamma (1+i\eta
_{23})\times $ $_{1}F_{1}(-i\eta _{23},1;i[kr-\mathbf{k}\cdot \mathbf{r}])$,
where $_{1}F_{1}$ is the confluent hypergeometric function, $\Gamma $ is the
Gamma function and $\eta _{23}$ is the Sommerfeld parameter, furthermore%
\begin{equation}
E_{rel}=\frac{\hbar ^{2}\mathbf{k}^{2}}{2m_{0}a_{23}}  \label{Erel}
\end{equation}%
\ and 
\begin{equation}
E_{CM}=\frac{\hbar ^{2}\mathbf{K}^{2}}{2m_{0}\left( A_{2}+A_{3}\right) }.
\label{ECM}
\end{equation}

The two important limits of the eigenstate $\left\vert 2,3\right\rangle
_{sc} $ in the case of screened Coulomb potential \ are the 
solution in the nuclear volume and the solution in the screened regime. (In
the screened case the coordinate representation $\langle \mathbf{R,r}%
\left\vert 2,3\right\rangle _{sc}$ is denoted by $\varphi _{23}\left( 
\mathbf{R,r}\right) _{sc}$.)

In the screened case and in the nuclear volume the approximate form $\left( %
\ref{Cbapp}\right) $ $\left( \varphi _{Cb,a}(\mathbf{r})=e^{i\mathbf{k}\cdot 
\mathbf{r}}f_{23}(k)/\sqrt{V}\right) $ of the (unscreened) Coulomb solution $%
\left( \ref{Cb1}\right) $ may be used. Here $f_{23}(k)$ is the appropriate
Coulomb factor corresponding to particles $2$ and $3$. Thus%
\begin{equation}
\varphi _{23}\left( \mathbf{R,r,}nucl\right) _{sc}=\frac{e^{i\mathbf{KR}}}{%
\sqrt{V}}\varphi _{Cb,a}(\mathbf{k},\mathbf{r})  \label{fi-nucl}
\end{equation}%
is used in the range of the nucleus and in the calculation of the nuclear
matrix-element.

The other important limit of $\left\vert 2,3\right\rangle _{sc}$ is the
screened (outer) range where $\varphi _{23}\left( \mathbf{R,r}\right) _{sc}$
becomes%
\begin{equation}
\varphi _{23}\left( \mathbf{R,r,}out\right) _{sc}=\frac{e^{i\mathbf{KR}}e^{i%
\mathbf{kr}}}{V}  \label{fi-sc}
\end{equation}%
of energy eigenvalue also $E_{23}=E_{CM}+E_{rel}$. It is used in the
calculation of the Coulomb matrix element.

From the point of view of the processes investigated the initial state of
negligible wave number and energy $\left( E_{0}=E_{i}=0\right) $ can be
written as $\varphi _{i}=V^{-3/2}$ for particles $1$, $%
2$ and $3$ that are somewhere in the normalization volume. The intermediate
states of particles $2$ and $3$ are determined by the wave number vectors $%
\mathbf{K}$ and $\mathbf{k}$. In the case of the assisting particle $1$ the
intermediate state is a plane wave of wave number vector $\mathbf{k}_{1}$.

The matrix elements $V_{Cb,\nu i}$ of the screened Coulomb potential between
the initial and intermediate states are%
\begin{eqnarray}
V_{Cb}(1,s)_{\nu i} &=&\frac{z_{1}z_{s}}{2\pi ^{2}}e^{2}\frac{\left( 2\pi
\right) ^{9}}{V^{3}}\delta \left( \mathbf{k}_{1}+\mathbf{K}\right) \times 
\label{VCb1snui} \\
&&\times \frac{\delta \left( \mathbf{k}+a(s)\mathbf{k}_{1}\right) }{\mathbf{k%
}_{1}^{2}+q_{sc,1s}^{2}}\text{ \ \ \ \ }  \notag
\end{eqnarray}%
where%
\begin{equation*}
a(s)=\frac{-A_{3}\delta _{s,2}+A_{2}\delta _{s,3}}{A_{2}+A_{3}}\text{ and }%
s=2,3.
\end{equation*}%
which according to standard time independent perturbation theory
of quantum mechanics \cite{Landau} determine the first order change of the
wavefunction in the range $r\lesssim R_{0}$ ($R_{0}$ is the nuclear radius
of particle $3$) due to Coulomb perturbation as%
\begin{equation}
\delta \varphi \left( \mathbf{r}\right) =\sum_{s=2,3}\delta \varphi \left( s,%
\mathbf{r}\right)   \label{dfi}
\end{equation}%
\begin{eqnarray}
\delta \varphi \left( s,\mathbf{r}\right)  &=&\int \int \frac{%
V_{Cb}(1,s)_{\nu i}}{E_{\nu }-E_{i}}\frac{V}{\left( 2\pi \right) ^{6}}\times 
\label{dfis} \\
&&\times e^{i(\mathbf{KR}+\mathbf{\mathbf{k}_{1}\mathbf{r}_{1})}}\varphi
_{Cb,a}(\mathbf{k},\mathbf{r})d\mathbf{K}d\mathbf{k},  \notag
\end{eqnarray}%
where $E_{i}$ and $E_{\nu }$ are the kinetic energies in the initial and
intermediate states, respectively. The initial momenta and kinetic energies
of particles $1$, $2$ and $3$ are neglected $\left( E_{i}=0\right) $.%
\begin{equation}
E_{\nu }(\mathbf{K},\mathbf{k)}=\frac{\hbar ^{2}\mathbf{K}^{2}}{%
2m_{0}(A_{2}+A_{3})}+\frac{\hbar ^{2}\mathbf{k}^{2}}{2m_{0}a_{23}}+\frac{%
\hbar ^{2}\mathbf{k}_{1}^{2}}{2m_{0}A_{1}},  \label{Em}
\end{equation}%
Thus%
\begin{eqnarray}
\delta \varphi \left( s,\mathbf{r}\right)  &=&z_{1}z_{s}\alpha _{f}\frac{%
4\pi \hbar c}{V^{5/2}}\frac{e^{i(\mathbf{\mathbf{k}_{1}\mathbf{r}_{1}-k}_{1}%
\mathbf{R)}}}{\mathbf{k}_{1}^{2}+q_{sc,1s}^{2}}\times   \label{dfi2} \\
&&\times \frac{2m_{0}a_{1s}}{\hbar ^{2}\mathbf{k}_{1}^{2}}\left[
f_{23}\left( k\right) e^{i\mathbf{kr}}\right] _{\mathbf{k}=a(s)\mathbf{k}%
_{1}}.  \notag
\end{eqnarray}%
It can be seen that the arguments of $f_{23}\left( k\right) $ are $\frac{%
A_{3}}{A_{2}+A_{3}}k_{1}$ and $\frac{A_{2}}{A_{2}+A_{3}}k_{1}$, here $%
k_{1}=\left\vert \mathbf{k}_{1}\right\vert $. Consequently, if particle $1$
obtains large kinetic energy, as is the case in the nuclear reaction, then
the Coulomb factors $f_{23}\left( k\right) $ and the rate of the process too
will considerably increase. Furthermore, $\delta \varphi \left( \mathbf{r}\right) 
$, which causes the effect, is temperature independent. Up to this point the
calculation and the results are nuclear reaction and nuclear model
independent.

\subsection{Transition probability per unit time and cross section of $p$%
-capture}

Now we can calculate the rate of nuclear reaction due to the modification
caused by particle $1$. (The intermediate and final states of particle $1$
are identical.)

The Hamiltonian $V_{st}(2,3)$ of strong interaction which is responsible
for nuclear reactions between particles $2$ and $3$ is%
\begin{eqnarray}
V_{st}\left( 2,3\right) &=&-V_{0}\text{ \ if }\left\vert \mathbf{r}%
_{23}\right\vert =\left\vert \mathbf{r}\right\vert \leq b\text{ and}
\label{Vst1} \\
V_{st}\left( 2,3\right) &=&0\text{ \ if }\left\vert \mathbf{r}%
_{23}\right\vert =\left\vert \mathbf{r}\right\vert >b.  \notag
\end{eqnarray}%
We take $V_{0}=25$ $MeV$ and $b=2\times 10^{-13}$ $cm$ \cite{Blatt} in the
case of $pd$ reaction.

The final state of particle $1$ is a plane wave of wave number $\mathbf{k}%
_{1}$ and of kinetic energy $E_{1f}=\hbar ^{2}\mathbf{k}_{1}^{2}/\left(
2m_{1}\right) $. The final state of the captured proton has the form 
\begin{equation}
\varphi _{4}(\mathbf{R},\mathbf{r})=e^{i\mathbf{k}_{4}\cdot \mathbf{R}}\Phi
_{f}\left( \mathbf{r}\right) /\sqrt{V},  \label{fi4}
\end{equation}%
where $\Phi _{f}\left( \mathbf{r}\right) $ is the final nuclear state of the
proton in particle $4$ and $\mathbf{k}_{4}$ is the wave vector of particle $%
4 $. It has kinetic energy $E_{4f}=\hbar ^{2}\mathbf{k}_{4}^{2}/\left(
2m_{4}\right) $. The Weisskopf-approximation is used, i.e. $\Phi _{f}\left( 
\mathbf{r}\right) =\Phi _{fW}\left( \mathbf{r}\right) $ with

\begin{equation}
\Phi _{fW}\left( \mathbf{r}\right) =\sqrt{\frac{3}{4\pi R_{0}^{3}}}
\label{fi4W}
\end{equation}%
if $r\leq R_{0}$, where $R_{0}$ is the nuclear radius, and $\Phi _{fW}\left( 
\mathbf{r}\right) =0$ for $r>R_{0}$.

$V_{st,f\nu }$ is the matrix element of the potential of the strong
interaction between intermediate $\left( e^{i\mathbf{KR}}\varphi _{Cb,a}(%
\mathbf{k},\mathbf{r})/\sqrt{V}\right) $ and final $\left( e^{i\mathbf{k}%
_{4}\cdot \mathbf{R}}\Phi _{f}\left( \mathbf{r}\right) /\sqrt{V}\right) $
states.

Since $\Phi _{fW}\left( \mathbf{r}\right) $ and$\ V_{st}\left( \mathbf{r}%
\right) $ both have spherical symmetry the spherical term $\sin \left(
kr\right) /\left( kr\right) $ remains from $e^{i\mathbf{k}\cdot \mathbf{r}}$%
, which is present in $\varphi _{Cb,a}(\mathbf{k},\mathbf{r})$, in the
nuclear matrix-element. With the aid of the above wave function and the $b=R_{0}$
assumption 
\begin{equation}
V_{st,f\nu }^{W}=-V_{0}\frac{\sqrt{12\pi R_{0}}}{k}f_{23}(k)H\left( k\right) 
\frac{\left( 2\pi \right) ^{3}}{V^{3/2}}\delta \left( \mathbf{K}-\mathbf{k}%
_{4}\right)  \label{Vstf2'-2}
\end{equation}%
where 
\begin{equation}
H\left( k\right) =\int_{0}^{1}\sin (kR_{0}x)xdx.\text{ }  \label{I10}
\end{equation}%
According to standard time independent perturbation theory of quantum
mechanics the transition probability per unit time $\left(
W_{fi}^{(2)}\right) $ of the process can be written as%
\begin{equation}
W_{fi}^{(2)}=\frac{2\pi }{\hbar }\int \int \left\vert
T_{fi}^{(2)}\right\vert ^{2}\delta (E_{f}-\Delta )\frac{V^{2}}{\left( 2\pi
\right) ^{6}}d\mathbf{k}_{1}d\mathbf{k}_{4}  \label{Wfie}
\end{equation}%
with%
\begin{equation}
T_{fi}^{(2)}=\int \int \sum_{s=2,3}\frac{V_{st,f\nu }V_{Cb}(1,s)_{\nu i}}{%
E_{\nu }-E_{i}}\frac{V^{2}}{\left( 2\pi \right) ^{6}}d\mathbf{K}d\mathbf{k},
\label{Tif}
\end{equation}

Collecting everything obtained above, substituting $\left( \ref{Tif}\right) $
into $\left( \ref{Wfie}\right) $, neglecting $q_{sc,jk}^{2}$ since $%
q_{sc,jk}^{2}\ll \mathbf{k}_{1}^{2}=k_{0}^{2}$ with%
\begin{equation}
k_{0}=\hbar ^{-1}\sqrt{2m_{0}a_{14}\Delta }  \label{k0}
\end{equation}%
determined by energy conservation one can calculate $W_{fi}^{(2)}$. The
cross section $\sigma _{23}^{\left( 2\right) }$ of the process is defined as

\begin{equation}
\sigma _{23}^{\left( 2\right) }=\frac{N_{1}W_{fi}^{(2)}}{\frac{v_{23}}{V}}
\label{sigma23}
\end{equation}%
where $N_{1}$ is the number of particles $1$ in the normalization volume $V$
and $v_{23}/V$ is the flux of particle $2$ of relative velocity $v_{23}$.%
\begin{equation}
v_{23}\sigma _{23}^{\left( 2\right) }=n_{1}S_{pd}  \label{sigma23-2}
\end{equation}%
where $n_{1}=N_{1}/V$ is the number density of particles $1$ and%
\begin{eqnarray}
S_{pd} &=&24\pi ^{2}\sqrt{2}cR_{0}\frac{z_{1}^{2}\alpha
_{f}^{2}V_{0}^{2}\left( \hbar c\right) ^{4}}{\Delta ^{9/2}\left(
m_{0}c^{2}\right) ^{3/2}}\times  \label{SRANR} \\
&&\times \frac{\left( A_{2}+A_{3}\right) ^{2}}{a_{14}^{7/2}}\left[ F\left(
2\right) +F\left( 3\right) \right] ^{2}.  \notag
\end{eqnarray}%
with%
\begin{equation}
F(s)=\frac{z_{s}a_{1s}}{A_{3}\delta _{s,2}+A_{2}\delta _{s,3}}f_{23}\left[
a(s)k_{0}\right] H\left[ a(s)k_{0}\right] .  \label{Fs}
\end{equation}%
The magnitude of quantities $f_{23}\left[ a(s)k_{0}\right] $, $s=2,3$ mainly
determines the magnitude of the rate and power densities.

\subsection{Rate and power densities}

The rate $dN_{pd}/dt$ in the whole volume $V$ can be written as 
\begin{equation}
\frac{dN_{pd}}{dt}=N_{3}\Phi _{23}\sigma _{23}^{\left( 2\right) },
\label{dNfdt}
\end{equation}%
where $\Phi _{23}=n_{2}v_{23}$ is the flux of particles $2$ with $n_{2}$
their number density ($n_{2}=$ $N_{2}/V$) and $N_{3}$ is the number of
particles $3$ in the normalization volume. The rate density $%
r_{pd}=dn_{pd}/dt=$ $V^{-1}dN_{pd}/dt$ of the process can be written as%
\begin{equation}
r_{pd}=\frac{dn_{pd}}{dt}=n_{3}n_{2}n_{1}S_{pd},  \label{dnfdt}
\end{equation}%
where $n_{3}$ is the number density of particles $3$ ($n_{3}=N_{3}/V$). The
power density is defined as%
\begin{equation}
p_{pd}=\Delta \frac{dn_{pd}}{dt}=n_{1}n_{2}n_{3}P_{pd}  \label{powerdensity}
\end{equation}%
with $P_{pd}=S_{pd}\Delta $. The rate and power densities ($dn_{pd}/dt$ and $%
p_{pd}$) are temperature independent.

\subsection{Cross section of reactions with two final fragments in long
wavelength approximation}

In the case of reactions with two final fragments (see Fig.1b and $\left( %
\ref{Reaction 4}\right) $) the nuclear matrix element can be derived from the $%
S(E)$ quantity of $\left( \ref{sigma}\right) $, i.e. in long
wavelength approximation from $S(0)$ which is
the astrophysical factor at $E=0$. It can be done in the following manner.

Calculating the transition probability per unit time $W_{fi}^{(1)}$ of the
usual (first order) process in standard manner 
\begin{equation*}
W_{fi}^{(1)}=\int \frac{2\pi }{\hbar }\left\vert V_{st,fi}\right\vert
^{2}\delta \left( E_{f}-\Delta \right) \frac{V}{\left( 2\pi \right) ^{3}}d%
\mathbf{k}_{f}
\end{equation*}%
where $\mathbf{k}_{f}$ is the relative wave number of the two fragments of
rest masses $m_{4}=m_{0}A_{4}$, $m_{5}=m_{0}A_{5}$ and atomic numbers $A_{4}$%
, $A_{5}$, $E_{f}=$ $\hbar ^{2}\mathbf{k}_{f}^{2}/(2m_{0}a_{45})$ is the sum
of their kinetic energy and the nuclear matrix element is $V_{st,fi}$ having
the form $\left\vert V_{st,fi}\right\vert =f_{23}\left( k_{i}\right)
h_{fi}/V $ . Here $f_{23}\left( k_{i}\right) $ is the Coulomb factor of the
initial particles $2$ and $3$ with $k_{i}$ the magnitude of their relative
wave number vector $\mathbf{k}_{i}$. (The Coulomb factor $f_{45}\left(
k_{f}\right) \approx 1$ of the final particles $4$ and $5$ with $k_{f}$ the
magnitude of their relative wave number vector $\mathbf{k}_{f}$.) It is
supposed that $h_{fi}$ does not depend on $\mathbf{k}_{i}$ and $\mathbf{k}%
_{f}$ namely the long wavelength approximation is used. In this case the
product of the relative velocity $v_{23}$ of the initial particles $2$, $3$%
\emph{\ }and the cross section $\sigma _{23}$ is 
\begin{equation}
v_{23}\sigma _{23}^{\left( 1\right) }=\frac{\left\vert h_{fi}\right\vert ^{2}%
}{\pi \hbar }f_{23}^{2}\left( k_{i}\right) \frac{\left( m_{0}a_{45}\right)
^{3/2}}{\hbar ^{3}}\sqrt{2\Delta }.  \label{v23sigma23}
\end{equation}%
On the other hand, $v_{23}\sigma _{23}^{\left( 1\right) }$ is expressed with
the aid of $\left( \ref{sigma}\right) $ and $v_{23}=\sqrt{2E/\left(
m_{0}a_{23}\right) }$. From the equality of the two kinds of $v_{23}\sigma
_{23}^{\left( 1\right) }$ one gets%
\begin{equation}
\left\vert h_{fi}\right\vert ^{2}=\frac{\left( \hbar c\right) ^{4}S(0)}{%
z_{2}z_{3}\alpha _{f}\left( m_{0}c^{2}\right) ^{5/2}\sqrt{2\Delta }%
a_{45}^{3/2}a_{23}}.  \label{hfi2}
\end{equation}

In the case of the impurity assisted, second order process $\left\vert
V_{st,f\nu }\right\vert =f_{23}\left( k\right) \left\vert h_{fi}\right\vert
\left( 2\pi \right) ^{3}\delta \left( \mathbf{K}-\mathbf{K}_{f}\right) /V^{2}
$ where $\mathbf{K}_{f}$ and $\mathbf{k}_{f}$ are the final wave number vectors
attached to $CM$ and relative motions of the two final fragments, particles $%
4$ and $5$. $\mathbf{k}_{f}$ appears in $E_{f}$\ in the energy Dirac-delta.
Repeating the calculation of the transition probability per unit time of the
impurity assisted, second order process applying the above expression of $%
\left\vert V_{st,f\nu }\right\vert $ one gets%
\begin{equation}
v_{23}\sigma _{23}^{\left( 2\right) }=n_{1}S_{^{\prime }reaction^{\prime }},
\label{v23sigma23-2}
\end{equation}%
where%
\begin{equation}
S_{^{\prime }reaction^{\prime }}=\frac{8\alpha _{f}^{2}z_{1}^{2}}{%
a_{23}a_{123}^{3}}\frac{S(0)c}{m_{0}c^{2}}\left( \frac{\hbar c}{\Delta }%
\right) ^{3}I  \label{result2}
\end{equation}%
with%
\begin{equation}
I=\int_{0}^{1}\left( \sum_{s=2,3}\frac{z_{s}a_{1s}\sqrt{A_{s}}}{\sqrt{%
e^{b_{23}A_{s}\frac{1}{x}}-1}}\right) ^{2}\frac{\sqrt{1-x^{2}}}{x^{7}}dx.
\label{Iintcharged}
\end{equation}%
Here $b_{23}=2\pi z_{2}z_{3}\alpha _{f}\sqrt{m_{0}c^{2}/\left(
2a_{123}\Delta \right) }$ with $a_{123}=A_{1}\left( A_{2}+A_{3}\right)
/\left( A_{1}+A_{2}+A_{3}\right) $. In the index $^{\prime }reaction^{\prime
}$ the reaction resulting the two fragments will be marked.

\subsection{Cross section of reactions with two final fragments beyond long
wavelength approximation}

If $\left\vert h_{fi}\right\vert $ has $\mathbf{k}_{1}$ dependence then it
is manifested through the relative energy $\left( E\right) $ dependence of
the astrophysical factor $\left[ S(E)\right] $ and it can be expressed as%
\begin{equation}
\left\vert h_{fi}\left( k_{1}\right) \right\vert =\sqrt{\frac{\left( \hbar
c\right) ^{4}S\left[ E(k_{1})\right] }{z_{2}z_{3}\alpha _{f}\left(
m_{0}c^{2}\right) ^{5/2}\sqrt{2\Delta }a_{45}^{3/2}a_{23}}}  \label{hfiE}
\end{equation}%
where%
\begin{equation}
E(k_{1})=\left[ \frac{\hbar ^{2}\mathbf{k}^{2}}{2m_{0}a_{23}}\right] _{%
\mathbf{k}=a(s)\mathbf{k}_{1}}=\frac{\hbar ^{2}a^{2}(s)k_{1}^{2}}{%
2m_{0}a_{23}}.  \label{E-S(E)-ben}
\end{equation}%
Consequently%
\begin{equation}
S_{^{\prime }reaction^{\prime }}=\frac{8\alpha _{f}^{2}z_{1}^{2}}{%
a_{23}a_{123}^{3}}\frac{S(0)c}{m_{0}c^{2}}\left( \frac{\hbar c}{\Delta }%
\right) ^{3}J  \label{result 3}
\end{equation}%
with%
\begin{eqnarray}
J &=&\int_{0}^{1}\left( \sum_{s=2,3}\frac{z_{s}a_{1s}\sqrt{A_{s}}\sqrt{\frac{%
S\left[ E(s,x)\right] }{S(0)}}}{\sqrt{e^{b_{23}A_{s}\frac{1}{x}}-1}}\right)
^{2}  \label{Jint} \\
&&\times \frac{\sqrt{1-x^{2}}}{x^{7}}dx.  \notag
\end{eqnarray}%
The argument of $S(E)$ in the integrand is 
\begin{equation}
E(s,x)=\Delta \frac{a_{123}}{a_{23}}a^{2}(s)x^{2}.  \label{Esx}
\end{equation}

\subsection{Atomic atom-ionic gas mix and wall interaction}

It is plausible to extend the investigation to the possible consequence of
plasma-wall interaction. The role of particle $1$ is played by the wall which is
supposed to be a solid (metal) from atoms with nuclei of charge and mass
numbers $z_{1}$and $A_{1}$. For initial state a Bloch-function of the form%
\begin{equation}
\varphi _{\mathbf{k}_{1,i}}(\mathbf{r}_{1})=\frac{1}{\sqrt{N_{1}}}\sum_{%
\mathbf{L}}e^{i\mathbf{k}_{1,i}\cdot \mathbf{L}}a(\mathbf{r}_{1}-\mathbf{L}),
\label{Bloch}
\end{equation}%
is taken, that is localized around all of the lattice points  \cite{Ziman}.
Here $\mathbf{r}_{1}$ is the coordinate, $\mathbf{k}_{1,i}$ is wave number vector
of the first Brillouin zone ($BZ$) of the reciprocal lattice, $a(\mathbf{r}%
_{1}-\mathbf{L})$ is the Wannier-function, which is independent of $\mathbf{k}%
_{1,i}$ within the $BZ$ and is well localized around lattice site $\mathbf{L}
$. $N_{1}$ is the number of lattice points of the lattice of particles $1$.
Repeating the transition probability per unit time and cross section
calculation applying $\left( \ref{Bloch}\right) $ (after a lengthy
calculation which is omitted here) it is obtained that cross section results
(formulae $\left( \ref{sigma23-2}\right) $, $\left( \ref{SRANR}\right) $ and 
$\left( \ref{Fs}\right) $ in case of proton capture, $\left( \ref{result2}%
\right) $, $\left( \ref{Iintcharged}\right) $ and $\left( \ref{result 3}%
\right) $, $\left( \ref{Jint}\right) $ in case of reactions with two final
fragments) remain unchanged and $n_{1}=N_{1c}/v_{c}$, where $v_{c}$ is the
volume of elementary cell of the solid and $N_{1c}$ is the number of
particles $1$ in the elementary cell.

\section{Rate and power densities in a $p-d-Xe$ atomic atom-ionic gas mix}

Reaction 
\begin{equation}
p+d\rightarrow _{2}^{3}He+\gamma +5.493MeV  \label{normal-pd}
\end{equation}%
is not suitable for energy production since its cross section (the $S(0)$
value, see \cite{Angulo}) is rather small compared with other candidate
reactions and only a minor part of the reaction energy $\Delta =5.493$ $%
MeV=8.800\times 10^{-13}$ $J$ is taken away by $^{3}He$ ($E_{^{3}He}=\Delta
^{2}/\left( 6m_{0}c^{2}\right) =5.4$ $keV$) and the main part $E_{\gamma
}=5.488$ $MeV$ is taken away by $\gamma $ radiation which is difficult to
convert to heat. However, in reaction $\left( \ref{Reaction 2}\right) $ the
reaction energy is taken away by particles $_{2}^{3}He$ and $%
_{z_{1}}^{A_{1}}V^{\prime }$ as their kinetic energy that they can lose in a
very short range to their environment converting the reaction energy
efficiently into heat. Therefore the direct observation of $_{2}^{3}He$ and $%
_{z_{1}}^{A_{1}}V^{\prime }$ is hard.

The rate $\left( dn_{pd}/dt\right) $ and power $\left( \Delta
dn_{pd}/dt\right) $ densities of impurity assisted $p+d\rightarrow $ $%
_{2}^{3}He$ reaction are determined by $\left( \ref{dnfdt}\right) $ and $%
\left( \ref{powerdensity}\right) $ with 
\begin{equation}
S_{pd}=1.89\times 10^{-53}z_{1}^{2}\text{ }cm^{6}s^{-1},  \label{S}
\end{equation}%
where $z_{1}$ is the charge number of the assisting nucleus and with 
\begin{equation}
P_{pd}=1.66\times 10^{-65}z_{1}^{2}\text{ }cm^{6}W,  \label{P}
\end{equation}%
respectively. Taking $z_{1}=54$ ($Xe$) and $n_{1}=n_{2}=n_{3}=2.65\times
10^{20}$ $cm^{-3}$ ($n_{1}$, $n_{2}$ and $n_{3}$ are the number densities of 
$Xe$, $p$ and $d$, i.e. particles 1, 2 and 3) one gets for rate and power densities considerable values:
\begin{equation}
r_{pd}=dn_{pd}/dt=1.02\times 10^{12}\text{ }cm^{-3}s^{-1}
\label{rate density}
\end{equation}%
and 
\begin{equation}
p_{pd}=\Delta dn_{pd}/dt=0.901\text{ }Wcm^{-3}  \label{power density}
\end{equation}.
If the impurity is $Hg$ or $U
$ then the above numbers must be multiplied by $2.2$ or $2.9$, respectively.

One must emphasize that both rate and power densities ($dn_{pd}/dt$ and $%
p_{pd}$) are temperature independent. It must be mentioned too that the effect
is not affected by Coulomb screening and the only condition is that
the participants must be in atomic or in atom-ionic state. This requirement
and the temperature independence of $dn_{pd}/dt$ and $p_{pd}$ greatly weaken
the necessary conditions.

\section{Other impurity assisted nuclear reactions}

Now let us consider the impurity assisted proton captures $\left( \ref%
{Reaction 1}\right) $ in general. The reaction energy $\Delta $ is the
difference between the sum of the initial and final mass excesses, i.e. $%
\Delta =\Delta _{p}+\Delta _{A_{3},z_{3}}-\Delta _{A_{3}+1,z_{3}+1}$. Since
particle $1$ assists the nuclear reaction its rest mass does not change. $%
\Delta _{p}$, $\Delta _{A_{3},z_{3}}$ and $\Delta _{A_{3}+1,z_{3}+1}$ are
mass excesses of proton, $_{z_{3}}^{A_{3}}X$ and $_{z_{3}+1}^{A_{3}+1}Y$
nuclei, respectively \cite{Shir}. Moreover, the capture reaction may be
extended to the impurity assisted capture of particles $_{z_{2}}^{A_{2}}w$
(see reaction $\left( \ref{Reaction 3}\right) $), e.g. the capture of
deuteron $\left( d\right) $, triton $\left( t\right) $, $^{3}He$, $^{4}He$,
etc.. In this case $\Delta =\Delta _{A_{2},z_{2}}+\Delta
_{A_{3},z_{3}}-\Delta _{A_{3}+A_{2},z_{3}+z_{2}}$. $\Delta _{A_{2},z_{2}}$, $%
\Delta _{A_{3},z_{3}}$ and $\Delta _{A_{3}+A_{2},z_{3}+z_{2}}$ are the
corresponding mass excesses. The mechanism discovered makes also possible
reaction $\left( \ref{Reaction 4}\right) $ with conditions $%
A_{2}+A_{3}=A_{4}+A_{5}$ and $z_{2}+z_{3}=z_{4}+z_{5}$. In this case $\Delta
=\Delta _{A_{2},z_{2}}+\Delta _{A_{3},z_{3}}-\Delta _{A_{4},z_{4}}-\Delta
_{A_{5},z_{5}}$ where $\Delta _{A_{j},z_{j}}$ are the corresponding mass
excesses. Investigating the mass excess data \cite{Shir} one can recognize
that in the case of processes $\left( \ref{Reaction 1}\right) $, $\left( \ref{Reaction
3}\right) $ and $\left( \ref{Reaction 4}\right) $ the number of
energetically allowed reactions is large, their usefulness from the point of
view of energy production is mainly determined by the magnitude of the
numerical values of the quantities $f_{23}$ belonging to the particular
reaction.

Impurity $\left( _{z_{1}}^{A_{1}}V\right) $ assisted $d-Li$ reactions may
take place with $_{3}^{6}Li$ and $_{3}^{7}Li$ isotopes:%
\begin{equation}
_{z_{1}}^{A_{1}}V+d+\text{ }_{3}^{6}Li\rightarrow \text{ }%
_{z_{1}}^{A_{1}}V^{\prime }+2_{2}^{4}He+22.372\text{ }MeV,  \label{Li5}
\end{equation}%
\begin{equation}
_{z_{1}}^{A_{1}}V+d+\text{ }_{3}^{7}Li\rightarrow _{z_{1}}^{A_{1}}V^{\prime
}+2_{2}^{4}He+n+15.122\text{ }MeV  \label{Li6}
\end{equation}%
and%
\begin{equation}
_{z_{1}}^{A_{1}}V+d+\text{ }_{3}^{7}Li\rightarrow _{z_{1}}^{A_{1}}V^{\prime
}+_{4}^{9}Be+16.696\text{ }MeV.  \label{Li7}
\end{equation}%
If there are deuterons present then the%
\begin{equation}
_{z_{1}}^{A_{1}}V+d+\text{ }_{z_{3}}^{A_{3}}X\rightarrow \text{ }%
_{z_{1}}^{A_{1}}V^{\prime }+\text{ }_{z_{3}+1}^{A_{3}+2}Y+\Delta 
\label{d-capture}
\end{equation}%
impurity assisted $d$ capture processes (see e.g. $\left( \ref{Li7}\right) $
and the $_{z_{1}}^{A_{1}}V+d+d\rightarrow $ $_{z_{1}}^{A_{1}}V^{\prime }+$ $%
_{2}^{4}He+23.847MeV$ reaction), furthermore the $_{z_{1}}^{A_{1}}V+d+d%
\rightarrow $ $_{z_{1}}^{A_{1}}V^{\prime }+n+$ $_{2}^{3}He+3.269$ $MeV$ , $%
_{z_{1}}^{A_{1}}V+d+d\rightarrow $ $_{z_{1}}^{A_{1}}V^{\prime }+p+t+4.033$ $%
MeV$ impurity assisted $dd$ reactions may also take place where the energy
of the reaction is carried by particles $_{z_{1}}^{A_{1}}V^{\prime }$, $%
_{z_{3}+1}^{A_{3}+2}Y$ and $_{z_{1}}^{A_{1}}V^{\prime }$, $_{2}^{4}He$,
which have momentum of equal magnitude but opposite direction, by particles $%
_{z_{1}}^{A_{1}}V^{\prime }$, $n$ and $_{2}^{3}He$ and by particles $%
_{z_{1}}^{A_{1}}V^{\prime }$, $p$ and $t$, respectively.

The results of $S_{^{\prime }reaction^{\prime }}$ and power density
calculations of some $Xe$ assisted reactions in long wavelength
approximation and with $n_{1}=n_{2}=n_{3}=2.65\times 10^{20}$ $cm^{-3}$ can
be found in Table I. From the point of view of rate and power densities the
screening of the Coulomb potential is not essential ($k_{0}\gg q_{sc}$)
consequently the above reactions bring up the possibility of a quite new
type of apparatus since the processes need atomic state of participant
materials only, i.e. need much lower temperature compared to the working
temperature of fusion power stations planned to date.

\begin{table}[tbp]
\tabskip=8pt 
\centerline {\vbox{\halign{\strut $#$\hfil&\hfil$#$\hfil&\hfil$#$
\hfil&\hfil$#$\hfil&\hfil$#$\hfil&\hfil$#$\cr
\noalign{\hrule\vskip2pt\hrule\vskip2pt}
Reaction& S(0)&S_{'Reaction'}&\Delta&p_{'Reaction'}\cr
\noalign{\vskip2pt\hrule\vskip2pt}
d(d,n)_{2}^{3}He & 0.055 & 1.01\times10^{-48} & 3.269 &9.82 \cr
d(d,p)t& 0.0571 & 1.10\times 10^{-48}&4.033& 13.2\cr
d(t,n)_{2}^{4}He & 11.7 & 1.06\times 10^{-46} &17.59& 5.57\times 10^{3} \cr
_{2}^{3}He(d,p)_{2}^{4}He & 5.9 & 1.51\times 10^{-48}&18.25& 82.6\cr
_{3}^{6}Li(p,\alpha)_{2}^{3}He & 2.97 & 1.99\times10^{-49} & 4.019 & 2.38 \cr
_{3}^{7}Li(p,\alpha )_{2}^{4}He & 0.0594 & 3.85\times10^{-51} & 17.347 & 0.199 \cr
_{4}^{9}Be(p,\alpha)_{3}^{6}Li & 17. & 1.79\times10^{-49} & 2.126 & 1.13 \cr
_{4}^{9}Be(p,d)_{4}^{8}Be & 17. & 1.66\times10^{-49} & 0.56 &0.277 \cr
_{4}^{9}Be(\alpha ,n)_{6}^{12}C & 2.5\times10^{3} & 6.22\times10^{-51} & 5.701 & 0.106 \cr
  & 6.\times10^{5} & 1.49\times10^{-48} & 5.701 & 25.4 \cr
_{5}^{10}B(p,\alpha )_{4}^{7}Be & 4 & 1.04 \times10^{-50} & 1.145 & 0.0356 \cr
  & 2.\times10^{3} & 5.21\times10^{-48} & 1.145 & 17.8 \cr
_{5}^{11}B(p,\alpha )_{4}^{8}Be & 187 & 5.16\times10^{-49} & 8.59 & 13.2 \cr
\noalign{\vskip2pt\hrule\vskip2pt\hrule}}}}
\caption{$S(0)$ is the astrophysical factor at $E=0$ in $MeVb$ \protect\cite%
{Angulo}, \protect\cite{Descou}. $S_{^{\prime }Reaction^{\prime }}$ (in $%
cm^{6}s^{-1}$) is calculated using $\left( \protect\ref{result2}\right) $
with $\left( \protect\ref{Iintcharged}\right) $ taking $z_{1}=54$ $\left(
Xe\right) $, $\Delta $ is the energy of the reaction in $MeV$ and $%
p_{^{\prime }Reaction^{\prime }}=\Delta n_{1}n_{2}n_{3}S_{^{\prime
}Reaction^{\prime }}$ is the power density in $Wcm^{-3}$ that is calculated
with $n_{1}=n_{2}=n_{3}=2.65\times 10^{20}$ $cm^{-3}$. In the case of $%
_{4}^{9}Be(\protect\alpha ,n)_{6}^{12}C$ and $_{5}^{10}B(p,\protect\alpha %
)_{4}^{7}Be$ reactions the astrophysical factor $[S(E)]$ has strong energy
dependence therefore the calculation was carried out with two characteristic
values of $S(E)$. }
\label{Table1}
\end{table}

If there is $Li$ present then%
\begin{equation}
_{z_{1}}^{A_{1}}V+\text{ }_{3}^{A_{2}}Li+\text{ }_{z_{3}}^{A_{3}}X%
\rightarrow \text{ }_{z_{1}}^{A_{1}}V^{\prime }+\text{ }%
_{z_{3}+3}^{A_{2}+A_{3}}Y+\Delta   \label{Li-capture}
\end{equation}%
impurity assisted $Li$ capture reactions may happen too. Let us examine the
impurity assisted%
\begin{equation}
_{z_{1}}^{A_{1}}V+\text{ }_{3}^{A_{2}}Li+\text{ }_{3}^{A_{3}}Li\rightarrow 
\text{ }_{z_{1}}^{A_{1}}V^{\prime }+\text{ }_{6}^{A_{2}+A_{3}}C+\Delta 
\label{PdLi}
\end{equation}%
$Li$ capture reactions and as an example let us take $z_{2}=z_{3}=3$, $%
A_{2}=6$, $A_{3}=7$, $A_{2}+A_{3}=A_{4}=13$, that corresponds to the%
\begin{equation}
_{z_{1}}^{A_{1}}V+\text{ }_{3}^{6}Li+\text{ }_{3}^{7}Li\rightarrow
_{z_{1}}^{A_{1}}V^{\prime }+\text{ }_{6}^{13}C+25.869\text{ }MeV
\label{PdLi2}
\end{equation}%
reaction. Taking $A_{1}=130$, $\eta _{23}(\frac{7}{13}k_{0})=0.487$, $\eta
_{23}(\frac{6}{13}k_{0})=0.568$\ and $f_{23}(\frac{7}{13}k_{0})=0.388$ and $%
f_{23}(\frac{6}{13}k_{0})=0.322$ (see $\left( \ref{etajk}\right) $, $\left( %
\ref{Fjk}\right) $, $\left( \ref{k0}\right) $ and $\left( \ref{SRANR}\right) 
$). These numbers are very promising. The reactions $_{z_{1}}^{A_{1}}V+$ $%
_{3}^{6}Li+$ $_{3}^{6}Li\rightarrow $ $_{z_{1}}^{A_{1}}V^{\prime }+$ $%
_{6}^{12}C+28.174$ $MeV$, $_{z_{1}}^{A_{1}}V+$ $_{3}^{6}Li+$ $%
_{3}^{6}Li\rightarrow $ $_{z_{1}}^{A_{1}}V^{\prime }+$ $3_{2}^{4}He+20.898$ $%
MeV$ and $_{z_{1}}^{A_{1}}V+$ $_{3}^{7}Li+$ $_{3}^{7}Li\rightarrow
_{z_{1}}^{A_{1}}V^{\prime }+$ $_{6}^{14}C+26.795$ $MeV$ may have importance
too. (The list is incomplete.)

\section{Summary}

The consequences of impurities in nuclear fusion fuels of plasma state are
discussed. According to calculations in certain cases second order processes
may produce greatly higher fusion rate than the rate due to direct (first
order) processes. In the examined problem it is found that Coulomb
scattering of the fusionable nuclei on the screened Coulomb potential of the
impurity can diminish the hindering Coulomb factor between them. Since the
second order process does not demand the matter to be in ionized state the
assistance of impurities can allow to decrease significantly the plasma
temperature which is determined only by the requirement that all components
must be in atomic or atom-ionic state. The results suggest that, on the
other hand, the density of the components has to be considerably increased.
The effective influence of wall-gas mix interaction brings up the possible
importance of gas mix-metal surface processes too. Promising new fuel mixes
are also put forward. Based of these results it may be expected that
search for new approach to energy production by nuclear fusion may be
started.

\end{document}